\documentclass[aps,prl,preprint,superscriptaddress]{revtex4}
\usepackage{amsmath}
\usepackage{gensymb}
\usepackage{subfig}
\usepackage{graphicx}
\usepackage{float}
\usepackage{amsthm}
\usepackage{amssymb}
\usepackage{xfrac}
\usepackage[toc,page]{appendix}
\usepackage{setspace}
\usepackage{lscape}
\usepackage[utf8]{inputenc}
\usepackage{dirtytalk}
\usepackage{tikz}
\newcommand{\hb}{\hbar}

\newcommand{\gm}{\gamma}
\newcommand{\lb}{\lambda}

\newcommand{\tbf}{\textbf}
\newcommand{\eps}{\epsilon}

\newcommand{\non}{\nonumber}

\newcommand{\og}{\omega}

\newcommand{\Gm}{\Gamma}

\newcommand{\sbp}{\subparagraph*{}}
\newcommand{\comment}[1]{}
\newcommand{\Tr}{{\rm Tr}}

\newcommand{\bds}{\boldsymbol}

\newcommand{\rgl}{\rangle}
\newcommand{\lgl}{\langle}

\newcommand{\dnw}{\downarrow}
\newcommand{\Dlt}{\Delta}

\newcommand{\alld}{\allowdisplaybreaks}

\newcommand{\expvc}[1]{\left\lgl  #1 \right\rgl}

\newcommand{\eqtion}[1]{\begin{equation} \non #1 \end{equation}}

\newtheorem{theorem}{Theorem}

\usepackage{hyperref}
\begin{document}
\title{Quantum Secrecy in Thermal States II}% : \\ Making our secrets secreter} % \\ Some more magic in microwave?  \date{}}
\author{Elizabeth Newton}%, Anne Ghesqui\`ere, Freya L. Wilson, Benjamin T. H. Varcoe and Martin Moseley}
\author{Anne Ghesqui\`ere}
\author{Freya L. Wilson}
\author{Benjamin T. H. Varcoe}
\affiliation{Quantum Experimental Group, School of Physics and Astronomy, University of Leeds, Leeds LS2 9JT, United Kingdom}
\email[]{a.ghesquiere@leeds.ac.uk}

\author{Martin Moseley}
\affiliation{Airbus Defense \& Space}
\date{\today}

\begin{abstract}
In this paper we consider a scheme for cryptographic key distribution based on a variation of continuous variable quantum key distribution called central broadcast. 
In the continuous variable central broadcast scheme, security arises from discord present in the Hanbury Brown and Twiss effect from a thermal source.
The benefit of this scheme is that it expands the range of frequencies into the microwave regime.
Longer wavelengths, where the thermal photon number is higher and correlations remain robust over long distances, may even be preferable to optical wavelengths. 
Assming that Alice controls the source but not the distribution of the light (eg satellite broadcasts), then we demonstrate that the central broadcast scheme is robust to an entangling cloner attack. 
We establish the security of the protocol both experimentally and theoretically.
\end{abstract}

\maketitle

Quantum key distribution (QKD) is rapidly gaining widespread acceptance \cite{Swiss:2007} as a method of secure key exchange and several high bandwidth devices have been demonstrated.% \cite{some more tech}. 
However, having distributed information across a network, there remains a limitation of key exchange at the user access point. 
For the end user, wireless access is the ideal use model.
The user access system must be both inexpensive and accessible without compromising security and maintaining the ability to work on scales of the order of metres to tens of metres. 

%\section{Motivation}

Recently, the potential of thermal states for QKD has been established \cite{sakuya:2017_2, Qi:2018}.
Although thermal states have sometimes been described as too noisy \cite{weedbrookprl:2010, weedbrookpra:2012}, they exhibit Hanbury Brown and Twiss correlations which have been found to exhibit positive discord \cite{ragy:2013}, a necessary condition for QKD\cite{pirandola:2014}.  

%In recent works \cite{sakuya:2017_2, Qi:2018}, the potential of thermal states for QKD has been established. 
%Although long decried as too noisy, especially in microwaves, these states have the advantage that the photons which make it travel in correlated pairs.
%This is known as the Hanbury Brown and Twiss effect; it has been explained both semi-classically and quantum mechanically \cite{Purcell:1956, Mandel:1958, Mandel:1959, Loudon:2000, Fox:2006, Mandel:1995}. 
%These correlations have been found to exhibit positive discord \cite{Ragy:2013}, a necessary condition for QKD \cite{Pirandola:2014}.
Consider a central broadcast protocol in which the radiation is split between twol parties, who now have correlated signals from which they can build a key. 
Another advantage to using thermal states is that they are easy and low-cost to produce.
Whereas large-scale implementations of QKD such as those described above require specific infrastructure, thermal states central broadcasting protocols can be implemented over short distances, with low-power devices. 

In the scheme proposed in \cite{sakuya:2017_2}: a thermal source is incident on a beamsplitter, with one output port connected to Alice and the other to Bob.  
We assumed that Alice controls the source, the channel leading to the beamsplitter and the beamsplitter itself.
They also control the channel separating them from the beamsplitter.
The only part opened to Eve resides on the branch between the beamsplitter and Bob.
We found that there is both a positive key rate and positive discord between the legal parties, both at optical frequencies (experimental result) and microwave frequencies (theoretical analysis).
In \cite{Qi:2018}, the authors shine a thermal source on a beamsplitter to prepare the states used by the legal parties in QKD, and find that such a source average photon number of 100 allows for efficient passive QKD.

In this paper, we relax our security by surrendering control of the channel between the source and the beamsplitter, leaving it open to attack.
Alice retains control of the source, the beamsplitter, their channel and detector. 
Eve can then attack the channel going to Bob (\cite{sakuya:2017_2}) or the channel going to the beamsplitter. 
This aims to provide Eve greater knowledge of the states making up the thermal radiation.

In the following, we describe the protocol and its modelling in more details.
%Musing on eavesdropper detection guides our choice of secrecy witness.
Finally we present theoretical and experimental results that demonstrate the security of this scheme.

\section{Protocol}
\begin{figure}
     \centering
     \includegraphics[scale=0.6]{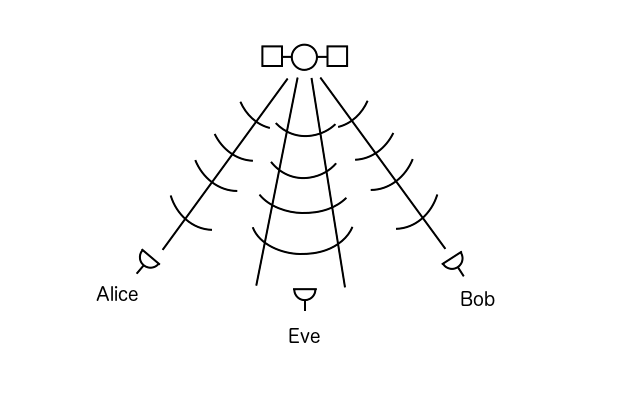}
          \caption{In this situation, a satellite beams down a signal, which is received by Alice, Bob and Eve. Eve can have a very large portion of the signal, but she does not control the signal being emitted.}
     \label{satellite_setup}
\end{figure}

\sbp This protocol is illustrated on Figure~\ref{satellite_setup}. 
A source (for instance a trusted satellite) emits thermal radiation which is picked by the legal parties and the eavesdropper.
We can consider that Eve can access quite a large portion of the signal, intercepting much of what should go to Alice and Bob.
We model this by giving Eve an entangling cloner, so she can divert as much of the signal to her as convenient.
However, we consider that the source is trusted; this means that the eavesdropper does not use the satellite to relay her own signal. 

We express the protocol formally as follows :

\begin{itemize}
\item Alice creates a beam from a trusted thermal source. 
\item On the way to their trusted beamsplitter with transmittance $\eta_2$, the signal is interfered with by Eve, via an entangling cloner denoted $\eta_1$.
\item Alice uses  $\eta_2$ to divert part of the signal to her detector and send the rest on to Bob. 
\item Similarly to \cite{sakuya:2017_2}, the bunched nature of the pairs coming out of $\eta_1$ means that fluctuations present at Alice's detector are correlated with those at Bob's detector.
\item To derive their data, Alice and Bob slice these fluctuations as convenient; as an example, a fluctuation above the mean could be a 1 and a fluctuation below the mean, a 0. 
\item Like any QKD scheme, our protocol requires quantum correlations. To confirm that the signal from Alice and Bob are correlated is done through verifying the thermal nature of their signal. Thus, Alice sends Bob small chunks of data for him to perform a $g^{(2)}$ calculation. A $g^{(2)} > 1$ means that the signal is thermal.
\item Alice and Bob now have a stream of independent and randomly correlated bits from which they can derive a key, the security of which they can improve with Cascade and Advantage Distillation, as per any QKD scheme.
\end{itemize}

\begin{figure}
     \centering
     \includegraphics[scale=0.9]{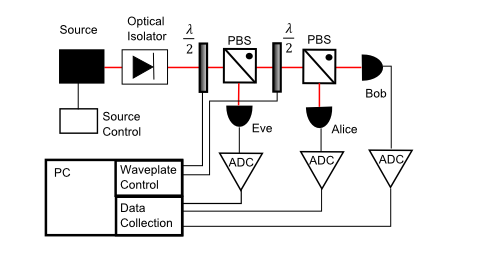}
          \caption{This is the schematic for this iteration of the protocol. The change in slight : effectively, Eve and Alice switch places, hence the deceiving resemblance between the schematics as shown in \cite{sakuya:2017_2} and this one. }
     \label{protocol_setup}
\end{figure}

This scheme was implemented as shown on Figure~\ref{protocol_setup}.
In order to simulate high levels of noise, we consider two attenuator channels between $\eta_2$ and  the legal parties, equivalent to adding a beamsplitter of transmittance $\eta_3$ between $\eta_2$ and Alice ($\eta_4$ for Bob) , with a input state of variance $N_3$ (and $N_4$) at the second input arm.

Once again, this is not a prepare-and-send scheme.
Alice controls the source, but the process of splitting pairs happening at the beamsplitter is stochastic, therefore unpredictable. 
Eve has no access to the channels between $\eta_2$ and either Alice or Bob, nor any control over their detectors.

\subsection{Theoretical modelling}

\sbp Let us recall that thermal states can be modelled using Gaussian statistics, which makes them easily defined and manipulated through their first and second moments \cite{Eisert:2003, GarciaPatron:2007}. 
The former are contained in the displacement vector $\expvc{\hat{r}}$, where $\hat{r}$ is the system's operator, and $\rho$ the state's density operator. 
The second moments are contained in the covariance matrix $\gm$ defined as \eqtion{\gm_{ij} = \Tr \left[ \rho \left\{ (\hat{r}_i - \expvc{\hat{r}_i}),(\hat{r}_j - \expvc{\hat{r}_j})\right\}\rho \right]\,,} where we write the anti-commutator using $\left\{ \right\}$. 
A thermal state has covariance matrix $\gm_{in} = 2(\bar{n} + 1) \bds{I} $, where $\bar{n}$ is the average photon number and $\bds{I}$ the identity matrix, and null displacement.
%\textit{We consider in the present work, a displaced thermal state, with covariance matrix as before, but with non-null displacement. }
We use the Bose Einstein distribution \begin{equation} \label{BEdist} \bar{n} = \frac{1}{e^{\sfrac{\hb \og}{k_B T}}-1}\,,\end{equation} and consider \comment{\textbf{narrowband???}} detectors measuring radiation at $30GHz$ and $T=300K$, so that $\bar{n} = 1309$. 

The beamsplitters are modelled as \eqtion{\bds{V}_i = \left( \begin{array}{cc} \sqrt{\eta_i} \bds{I} & \mu_i \bds{I} \\ -\mu_i \bds{I} & \sqrt{\eta_i} \bds{I} \end{array} \right) \,,}
where $\eta_i$ is the transmittance and $\mu_i = \sqrt{1 - \eta_i}$ represents the noise. 
They act on the state as $\gm_{out} = \bds{V} \gm_{in} \bds{V}^T$.

The input state at the first beamsplitter contains the thermal source and Eve's source; it has covariance matrix and displacement vector
\eqtion{\gm_{in} = \left(\begin{array}{cccc} V_s^x & 0 & 0 & 0\\ 0 & V_s^p & 0 & 0 \\ 0 & 0 & V_e^x & 0 \\ 0 & 0 & 0 & V_e^p \end{array} \right) \qquad \tbf{r}_{in} = (x_s; p_s; x_e; p_e)^T  \, .}
We note the structure of the covariance matrix as $\gm_{in} = \gm_{source} \bigoplus \gm_{Eve}$. The two empty sub-matrices would represent potential pre-existing correlations between the source and Eve, which in our set-up, is unrealistic.

The output of the second beamsplitter is 
\eqtion{\gm_{out} = \left(\begin{array}{ccc} \tilde{\gm_b} & \tilde{\gm_{ab}} & \tilde{\gm_{eb}} \\ \tilde{\gm_{ab}} & \tilde{\gm_a} & \tilde{\gm_{ea}} \\ \tilde{\gm_{eb}} & \tilde{\gm_{ea}} & \tilde{\gm_e} \end{array}\right) \,.}

We make the channel between $\eta_2$ and Alice, and between $\eta_2$ and Bob thermal noise channels by inputting states of variance $N_3$ on Alice's branch and $N_4$ on Bob's as
\eqtion{N_i = \frac{\eta_i \chi_i}{1-\eta_i} \,, \qquad \text{with} \qquad\chi_i = \frac{1-\eta_i}{\eta_i} + \eps_i \,,} and $\eps_i$ the channel excess noise \cite{GarciaPatron:2007}. 
The input state at $\eta_3$ and $\eta_4$ is then \eqtion{\gm_{int} = \left( \begin{array}{cc} N_3 & 0 \\ 0 & N_3 \end{array}\right) \bigoplus \gm_{out} \bigoplus \left( \begin{array}{cc} N_4 & 0 \\ 0 & N_4 \end{array}\right) \,,}
where $\gm_{out}$ is the state at the output of $\eta_2$, the first block sub-matrix is the input state at $\eta_3$ and the last sub-matrix, the input state at $\eta_4$.

The output covariance matrix is
\eqtion{\Gm_{out}= \left(\begin{array}{ccccc} \tilde{\Gm_v}  & \tilde{\Gm_{va}} & \tilde{\Gm_{ve}} & \tilde{\Gm_{vb}}  & \tilde{\Gm_{vv'}} \\ \tilde{\Gm_{va}} & \tilde{\Gm_{a}} & \tilde{\Gm_{ea}} & \tilde{\Gm_{ab}}  & \tilde{\Gm_{av'}} \\ \tilde{\Gm_{ve}}  & \tilde{\Gm_{ea}}  & \tilde{\Gm_e}  & \tilde{\Gm_{eb}}  & \tilde{\Gm_{ev'}}   \\ \tilde{\Gm_{vb}}  & \tilde{\Gm_{ab}} & \tilde{\Gm_{eb}}   & \tilde{\Gm_b}  & \tilde{\Gm_{bv'}} \\ \tilde{\Gm_{vv'}}  & \tilde{\Gm_{av'}} & \tilde{\Gm_{ev'}}  & \tilde{\Gm_{bv'}}  & \tilde{\Gm_{v'}}  \end{array}\right)} 
where the block sub-matrices are given in the appendix. 

\sbp Maurer and Wolf \cite{Maurer:1999}  have proved a theorem providing conditions to be satisfied for a scheme such as ours to be secure. 
The theorem reads as follows : 

[quote]
\begin{theorem} In Scenario 1, the following conditions are equivalent :
\begin{enumerate}\item $I(A:B|E) >0$ \item $K(A:B \parallel E) >0$ \item $I(A:B\dnw E) >0$ \end{enumerate}
\end{theorem} [end quote] \footnote{even though this is a direct quote, we have adapted the notation to our scheme.}

where $K(A:B \parallel E)$ is the secret key rate. 
The third condition is actually the most restrictive. 
$I(A:B \dnw E)$ is the intrinsic conditional mutual information; it determines the unreducible amount of conditional mutual information between Alice and Bob, regardless of any attemps by Eve at acquiring more information through local operations; in other words, it is information inaccessible to Eve. 
Furthermore, it satisfies \eqtion{I(A:B\dnw E) < I(A:B |E) \,,} which makes it a tighter condition on the secret key rate.  

We can see its relation to the quantum discord if we recall that the latter, $D(B|A)$, is defined as the difference between the mutual information $I(A:B)$ and the classical mutual information $J(B|A)$ (or $J(A|B)$). $I(A:B)$ quantifies all possible correlations between Alice and Bob, but $J(B|A)$ quantifies those measured by local operations at Alice's and Bob's sites. 
Therefore, it can be understood as the intrinsic conditional mutual information as described previously. 
Let us therefore, rewrite the theorem as :
\begin{theorem} In our central broadcast scheme, the following conditions are equivalent:
\begin{enumerate} \item \label{CMI} $I(A:B|E) >0$ \item \label{SKR} $K(A:B \parallel E) >0$ \item \label{QD} $D(B|A) >0$ \end{enumerate}
\end{theorem}

It is therefore enough in principle, to demonstrate that either condition is satisfied. 
We shall however, prove two, namely the positivity of the conditional mutual information and that of the discord. 
The latter will allow us to demonstrate the quantum nature of the secrecy.

\sbp The mutual information $I(A:B)$ is given by \eqtion{I(A:B) = S(\Gm_a) + S(\Gm_b) - S(\Gm_{ab})\,,} 
where $S(x)$ is the Von Neumann entropy and $\Gm_{i}$ the covariance matrices of A, B and AB respectively. 
The Von Neumann entropy is given by \eqtion{S(x) = \sum_{i=1}^N \left(\frac{x_i+1}{2}\right)\log\left(\frac{x_i+1}{2}\right) - \left(\frac{x_i-1}{2}\right)\log\left(\frac{x_i-1}{2}\right)} 
where $x_i$ are the symplectic eigenvalues of $\Gm$.
The discord is defined explicitely as \eqtion{D(B|A) = S(\Gm_a) - S(\Gm_{ab}) +\min_{\Gm_0} S(\Gm_{b|x_A})\,}
where $\Gm_{b|x_A}$ is the covariance matrix of B conditionned by a homodyne measurement on A \cite{Weedbrookrmp:2012} \eqtion{\Gm_{b|x_A} = \, \Gm_b - \Gm_{ab}(X\Gm_a X)^{-1} \Gm_{ab}^T\,,} 
with $X = \left(\begin{array}{cc} 1& 0 \\ 0 & 0 \end{array}\right)$ and $()^{-1}$ the pseudo-inverse.
The conditional mutual information is \eqtion{I(A:B|E) = S(\Gm_{ae}) + S(\Gm_{be}) - S(\Gm_{e}) - S(\Gm_{abe})\,.}

%\section{Experimental results}
\section{Results and discussion}
\sbp The protocol was realised experimentally.
The thermal source is provided by a superluminescent diode coupled to an external cavity, making it a tuneable laser, run without any added modulation.
The laser can be run separately in coherent or in thermal mode, and the thermality of the source was established in \cite{sakuya:2017_2}.
The source bandwidth was measured at $\Dlt\lb = 0.4$nm spread around a centre wavelength of $\lb_0 = 780.09$nm; this give a   coherence time of $\tau_c = 4.8$ps.
The detectors are ThorLabs Det36A photodiodes, coupled to a LeCroy Waverunner 44xi oscilloscope; the combined integration time is 14ns and the oscilloscope samples at 5GSps.

The conditional mutual information is calculated from the sliced data strings using Shannon entropies $H(x) = -\sum p(x) \log(p(x))$ in terms of the measured frequencies $p(x)$.
%\textit{Since the set-up is very similar to that used in \cite{sakuya:2017_2}, the source is the same and its thermality has been established therein, we do not plot a $g^{(2)}$, to keep things concise.} 

\begin{figure}
    \centering
     \includegraphics[scale=0.4]{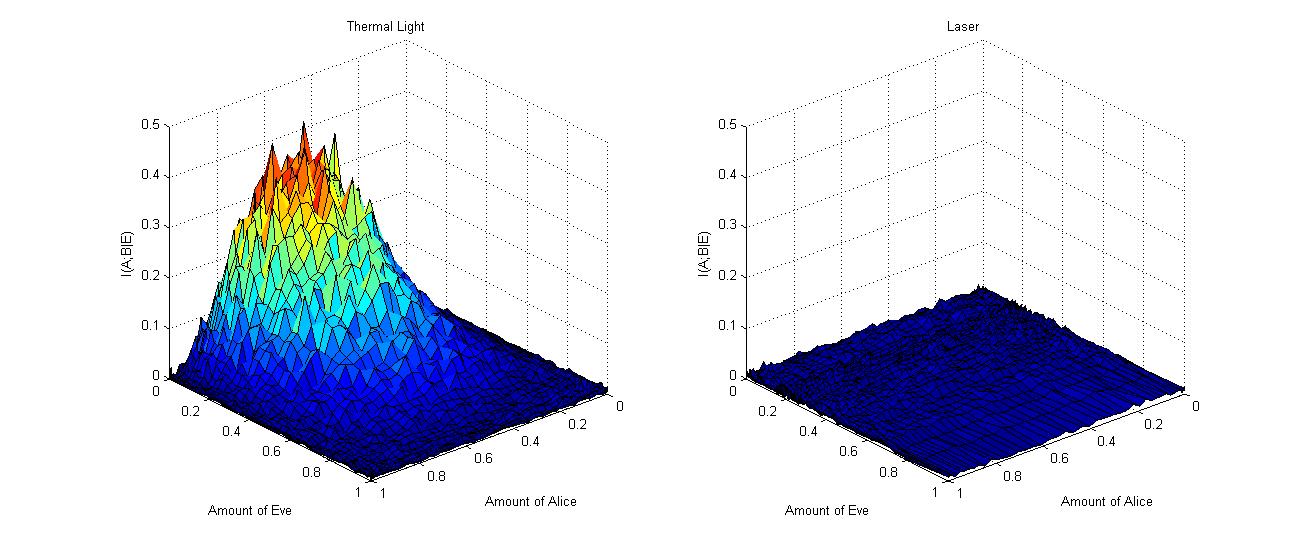} 
    \caption{Conditional mutual information for thermal states (left) versus coherent states (right). We can see that when $\eta_1 \rightarrow 1$, so when there is no amount of Eve coming between the legal parties, the conditional mutual information peaks.}
   \label{experimentalCMI}
\end{figure}

Figure~\ref{experimentalCMI} shows that the scheme works experimentally as predicted. 
$I(A:B|E)$ is best as $\eta_1$ tends to 1, and at $\eta_2 = 0.5$, so when Alice and Bob gets equal shares of most of the thermal source signal. 
This corresponds to a situation where the eavesdropper is absent, and where there is minimal loss.
As long as the $\eta_1 > 0.5$, the eavesdropper gets little of the signal and the advantage is to the legal parties.
However, no matter how much signal Eve receives, the conditional mutual information is always positive, and never exhibits a sharp fall-off, typical of point-to-point schemes over the $3dB$ limit.
This means that it is always possible to build key, albeit slowly.

Figure~\ref{experimentalCMI} allows us also to illustrate that this scheme cannot work in the coherent regime. 
As mentioned before in \cite{sakuya:2017_2}, coherent radiation is not bunched; therefore, it holds none of the intrinsic correlations contained in bunched pairs. 
There is no splitting of pairs occuring at the beamsplitters, because there are no such pairs; single photons travel through uncorrelated to Alice and Bob, who as a result can build no key from them.
This is shown on the right-hand graph of the figure.
$I(A:B|E)$ remains constant, no matter how much Eve lets through, no matter the split between Alice and Bob. 

%\section{Theoretical discussion}

\sbp Let us now compare these results to those obtained through our theoretical modelling. 
\begin{figure}
     \centering
\subfloat{\includegraphics[scale=0.28]{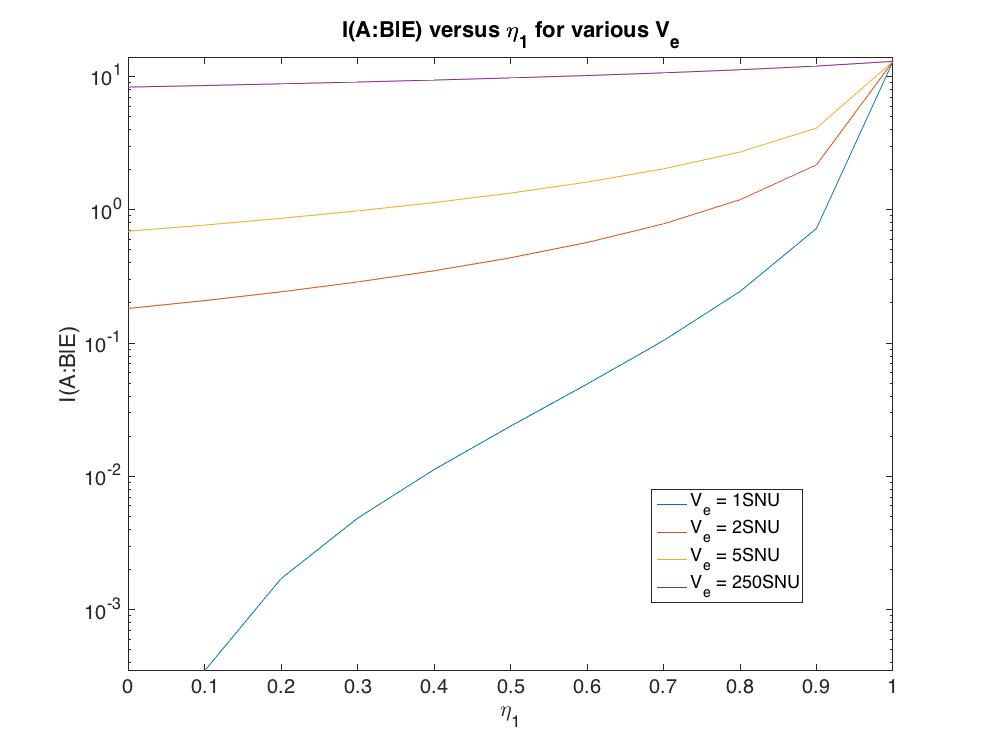} \label{Condmutinf}}
\subfloat{\includegraphics[scale=0.28]{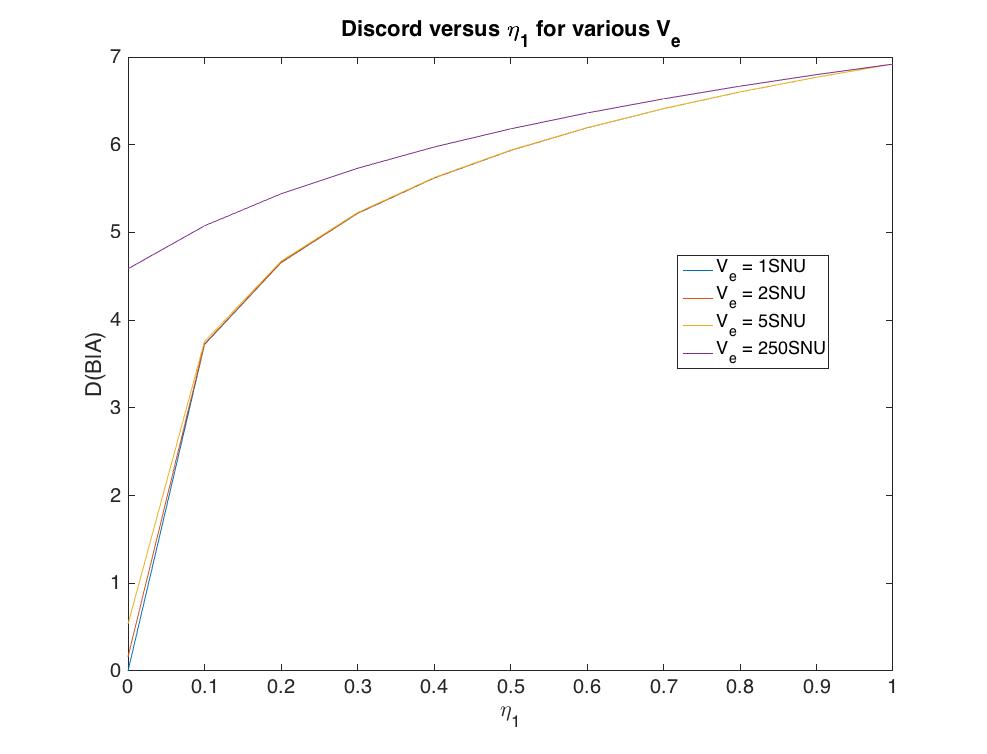} \label{Discords}}
     \caption{We plot the conditional mutual information $I(A:B|E)$ (left) and the discord $D(B|A)$ (right)against $\eta_1$, with $\eta_2 = 0.5$, $\eta_3 = \eta_4 = 0.2$ and $\eps_3 = \eps_4 = 10^{-2}$. At $\eta_2 = 0.5$, Alice and Bob share equal part of the signal. }
   %  \label{Condmutinf}
\end{figure} 

Figure~\ref{Condmutinf} shows the behaviour of the conditional mutual information as Eve lets more and more of the signal through. 
The plots match our experimental results. 
The higher $\eta_1$, the higher $I(A:B:|E)$. 
Also, since $I(A:B|E)$ is always positive, we conclude there always is secrecy in our scheme. 

We can also explore how the initial state of Eve influences the secrecy between Alice and Bob.
For that, we vary $V_e$ and see that as it increases, $I(A:B|E)$ increases also. 
The reason for this, we have mentioned before and will detail further in the lines below.

Figure~\ref{Discords} illustrates the positivity of the discord, regardless of $\eta_1$. 
This means that there always are quantum correlations between Alice and Bob. 
This satisfies the third of the conditions from our theorem, and we can affirm quantum secrecy. 

\sbp What is remarkable is the value of the discord when $\eta_1$ is null, so before Eve begins to let the source signal through.
In this case, what is actually measured is the amount of quantum correlations within Eve's state.
We have seen that the higher $V_e$ is, the higher $I(A:B|E)$, but here we see that the discord follows a similar trend.
This is particularly evident when $V_e = 250SNU$. 

This is a result of the physics of thermal states.
To understand this, let us step back and consider a single beamsplitter (input arms labelled 1 and 2, output arms labelled 3 and 4) with a thermal state at one input. 
Since it is bunched, there will be correlated photon pairs travelling into the beamsplitter. 
If both photons travel into the same input (say arm 1), we can expect three outputs \cite{Loudon:2000} : 
\begin{itemize} 
\item both photons are travelling through on arm 3 $P(2_3, 0_4)$, 
\item both photons travel onto arm 4 $P(0_3, 2_4)$ or 
\item one photon for each arm $P(1_3, 1_4)$. \end{itemize} 
This corresponds to Eve inputting a vacuum or a coherent state at arm 2 and why we can in fact equate her to any loss in the channel. 

On the other hand, if Eve inputs a thermal state as well, there is now a correlated pair of photons travelling into each input arm. 
This will gives us the following outputs: $P(4_3, 0_4)$, $P(0_3, 4_4)$, $P(2_3, 2_4)$, $P(3_3, 1_4)$,  and $P(1_3, 3_4)$. 
%\begin{itemize} \item $P(4_3, 0_4)$,  \item $P(0_3, 4_4)$, \item $P(2_3, 2_4)$, \item $P(3_3, 1_4)$, \item $P(1_3, 3_4)$. \end{itemize} 
The third case $P(2_3, 2_4)$, is three-degenerate; either both pairs get to the other side in one piece (which accounts for two degeneracies) or both pairs are split (the remaining degeneracy). 
This means that accounting for all possible outcomes, there are only two cases where there will not be at least one correlated pair travelling into $\eta_2$ to Alice and Bob: either one pair is split at $\eta_1$ and Eve gets three photons $P(3_3,1_4)$ (mitigated by the fact that Eve would choose to let most of the signal through at $\eta_1$ in order not to be noticed) or both pairs are split at $\eta_1$, which is one of the $P(2_3, 2_4)$ degeneracies.
%is only one case where there will not be at least one correlated pair travelling into $\eta_2$ to Alice and Bob. 

%This confirms what we have described previously.
If $V_e = 1SNU$, then Eve inputs a vacuum state, and Alice and Bob build key solely from the pairs produced at the source.
As a result, the discord is minimal at $\eta_1 \rightarrow 0$. 
If $V_e > 1SNU$, Eve's state can be regarded as thermal; in this case, she contributes pairs to those coming from the source.
In fact, if the eavesdropper's input is too significant, the legal parties can build a quantum secure key, regardless of how much signal is coming from the source. 
As in any QKD, we expect that the eavesdropper will try to minimise her input, if only to escape detection. 
At best, she can hope to merely ``listen'' in, in which case, her input is $V_e = 1SNU$. 
Yet, as soon as signal begins going through ($\eta_1>0.1$), the legal parties can build a quantum secure key, albeit slowly.

Let us point out that these plots have been obtained for very high level of noise on Alice's and Bob's branches.
Indeed $\eta_3$ and $\eta_4$ are such that $80\%$ of their signal is lost. 
Yet, even in this case, the legal parties are able to construct a quantum secure key.

\section{Concluding remarks}

\sbp 
In our previous protocol, the security arose from the quantum correlations within a pair which would split between Alice and Bob.
Since Eve placed herself on the arm going to Bob, she would interfere with/capture photons on their way to him, but she could not build a three way correlation sufficient to attack key exchange. 
In this paper, she places herself prior to the splitting of the pairs between Alice and Bob, interfering with the pairs directly from the source.
Unlike the situation in \cite{sakuya:2017_2}, she is not limited by the Heisenberg uncertainty principle and can intercept and resend bunched pairs at her leisure.
Therefore, the legal parties cannot distinguish the pairs coming from her to those coming from the source. 
This is not as bad as it sounds, however, because since Eve has no interaction with the output of $\eta_2$, any correlations within the pairs split at $\eta_2$ are completely safe from tampering. 
So actually, as long as Alice and Bob have a $g^{(2)}(0) > 1$, their pairs, regardless of their origins, are correlated, and quantum secrecy is possible \cite{sakuya:2017_2}.

This is the strength of this scheme.
Even if Eve succeeds in hiding in the noise, if her input is not either vacuum or a perfect coherent state, she will contribute correlations to the pool which Alice and Bob can build key from, but she cannot know when or if these injected states have contributed to the key. 
Another option for Eve is to actually become the source; we explore this in a forthcoming publication. 

This experiment was carried out at optical frequencies using a pseudo thermal source, however, the theoretical modelling was performed at values of $\bar{n}$ consistent with the microwave regime.
Interferometeres used in radio astronomy rely on the presence of thermal correlations being preserved over astronmical distances, and as the results in this paper suggest that the results are highly portable to the microwave regime.
Hence, this method of key exchange appears to be a viable option for long distance key exchange.

%%%%%%%%%%%%%%%%%%%%%%%%%%%%%%%%%%%%
\sbp The authors are grateful to network collaborators J. Rarity, S. Pirandola, C. Ottaviani, T. Spiller, N. Luktenhaus and W. Munro for very fruitful discussions. This work was supported by funding through the EPSRC Quantum Communications Hub EP/M013472/1 and additional funding for F.W. from Airbus Defense \& Space.  

\sbp Data that support the findings of this study are available from the Research Data Leeds Repository with the identifier \url{https://doi.org/10.5518/587} \cite{Data:QSTS2}.

\bibliographystyle{unsrt}

\bibliography{cvqkd2_bibli}

%%%%%%%%%%%%%%%%%%%%%%%%%%%%%%%%%%%%%%%%%%%%%%%%%%%%%%%%%%%%%%%%%%%%%%%%%%%%%%%%%%%%%%%%%%%%%

\appendix
\section{Protocol 2}
\subsection{After $\eta_2$}
The submatrices are as follows 
{\alld\begin{align} \tilde{\gm_b} =& \left(\begin{array}{cc}   \eta_2 + \mu_2 ^2(\eta_1 V_s^x + \mu_1^2 V_e^x) & 0 \\ 0 &  \eta_2 + \mu_2 ^2(\eta_1 V_s^p+ \mu_1^2 V_e^p)\end{array}\right) \non 
\\  \tilde{\gm_a} =&  \left(\begin{array}{cc} \mu_2 ^2 + \eta_2 (\eta_1 V_s^x + \mu_1^2 V_e^x ) & 0 \\ 0  & \mu_2 ^2 + \eta_2 (\eta_1 V_s^p + \mu_1^2 V_e^p) \end{array}\right) \non 
\\ \tilde{\gm_e} =& \left(\begin{array}{cc} \mu_1 ^2 V_s^x + \eta_1 V_e^x & 0 \\ 0 & \mu_1 ^2 V_s^p + \eta_1 V_e^p \end{array} \right) \non 
\\  \tilde{\gm_{ea}} =& \left( \begin{array}{cc} -\mu_1 \sqrt{\eta_1} \sqrt{\eta_2} (V_s^x-V_e^x) & 0 \\ 0  & -\mu_1 \sqrt{\eta_1} \sqrt{\eta_2} (V_s^p-V_e^p) \end{array} \right) \non 
\\ \tilde{\gm_{eb}} =& \left( \begin{array}{cc} -\mu_1 \sqrt{\eta_1} \mu_2 (V_s^x-V_e^x) & 0 \\ 0  & -\mu_1 \sqrt{\eta_1} \mu_2 (V_s^p-V_e^p) \end{array} \right) \non 
\\  \tilde{\gm_{ab}} =& \left(\begin{array}{cc} \mu_2 \sqrt{\eta_2} (\eta_1 V_s^x + \mu_1^2 V_e^x -1) & 0 \\ 0 & \mu_2 \sqrt{\eta_2} (\eta_1 V_s^p + \mu_1^2 V_e^p -1) \end{array}\right) \non
\end{align}}

\subsection{Ater $\eta_3$ and $\eta_4$}
The submatrices are as follows
{\alld\begin{align} \tilde{\Gm_e} =& \left(\begin{array}{cc} \expvc{\tilde{X_e}^2} & 0 \\ 0 & \expvc{\tilde{P_e}^2} \end{array} \right) \,, \quad \tilde{\Gm_a } = \left(\begin{array}{cc} \mu_3^2 N_3 + \eta_3 \expvc{\tilde{X_a}^2} & 0 \\ 0 & \mu_3^2 N_3 + \eta_3 \expvc{\tilde{P_a}^2}  \end{array} \right) \non \\
\tilde{\Gm_b}  =& \left(\begin{array}{cc} \mu_4^2 N_4 + \eta_4 \expvc{\tilde{X_b}^2} &  0 \\  0 & \mu_4^2 N_4 + \eta_4 \expvc{\tilde{P_b}^2}  \end{array} \right) \,, \quad \tilde{\Gm_v } = \left(\begin{array}{cc} \eta_3 N_3 + \mu_3^2 \expvc{\tilde{X_a}^2} & 0 \\ 0 & \eta_3 N_3 + \mu_3^2 \expvc{\tilde{P_a}^2}  \end{array} \right) \non \\ \non \\
\tilde{\Gm_{v'}} =& \left(\begin{array}{cc} \eta_4 N_4 + \mu_4^2 \expvc{\tilde{X_b}^2} & 0 \\ 0 & \eta_4 N_4 + \mu_4^2 \expvc{\tilde{P_b}^2}  \end{array} \right) \non \\
\tilde{\Gm_{ea} } =& \left(\begin{array}{cc} \sqrt{\eta_3} \expvc{\tilde{X_a} \tilde{X_e}} & 0 \\ 0 & \sqrt{\eta_3} \expvc{\tilde{P_a} \tilde{P_e}}  \end{array} \right) \, \quad 
\tilde{\Gm_{eb} }= \left(\begin{array}{cc} \sqrt{\eta_4} \expvc{\tilde{X_b} \tilde{X_e}} & 0 \\ 0 & \sqrt{\eta_4} \expvc{\tilde{P_b} \tilde{P_e}}  \end{array} \right) \non \\
\tilde{\Gm_{ab} } =& \left(\begin{array}{cc} \sqrt{\eta_3} \sqrt{\eta_4} \expvc{\tilde{X_a} \tilde{X_b}} & 0 \\ 0 & \sqrt{\eta_3} \sqrt{\eta_4}  \expvc{\tilde{P_a} \tilde{P_b}}  \end{array} \right) \,, \quad \tilde{\Gm_{vv'} }= \left(\begin{array}{cc} -\mu_3 \mu_4\expvc{\tilde{X_a} \tilde{X_b}} & 0 \\  & -\mu_3 \mu_4 \expvc{\tilde{P_a} \tilde{P_b}}  \end{array} \right) \non \\
\tilde{\Gm_{vb} } =& \left(\begin{array}{cc} \mu_3 \sqrt{\eta_4} \expvc{\tilde{X_a} \tilde{X_b}} & 0 \\ 0 & \mu_3 \sqrt{\eta_4}  \expvc{\tilde{P_a} \tilde{P_b}}  \end{array} \right) \,, \quad \tilde{\Gm_{av'}} = \left(\begin{array}{cc} -\sqrt{\eta_3}  \mu_4\expvc{\tilde{X_a} \tilde{X_b}} & 0 \\ 0 & -\sqrt{\eta_3}  \mu_4 \expvc{\tilde{P_a} \tilde{P_b}}  \end{array} \right) \non \\
\tilde{\Gm_{va} } =& \left(\begin{array}{cc} \mu_3 \sqrt{\eta_3} \left(\expvc{\tilde{X_a} ^2} - N_3 \right) & 0 \\ 0 & \mu_3 \sqrt{\eta_3}  \left(\expvc{\tilde{P_a}^2} - N_3 \right) \end{array} \right) \non \\
\tilde{\Gm_{bv'}} =& \left(\begin{array}{cc} \mu_4 \sqrt{\eta_4} \left(N_4  - \expvc{\tilde{X_b} ^2}\right) & 0 \\ 0  & \mu_4 \sqrt{\eta_4}  \left(N_4 - \expvc{\tilde{P_b}^2} \right) \end{array} \right) \non \\
\tilde{\Gm_{ve}} =& \left(\begin{array}{cc} \mu_3 \expvc{\tilde{X_a} \tilde{X_e}} & 0 \\ 0 & \mu_3 \expvc{\tilde{P_a} \tilde{P_e}}  \end{array} \right) \,, \quad
\tilde{\Gm_{ev'}} = \left(\begin{array}{cc} - \mu_4 \expvc{\tilde{X_b} \tilde{X_e}} & 0 \\ 0 & - \mu_4 \expvc{\tilde{P_b} \tilde{P_e}}  \end{array} \right) \non
\end{align}}

\end{document}